\documentclass[twoside]{dis04}
\def\runauthor{Marek Ta\v{s}evsk\'{y}}
\def\shorttitle{Forward Physics with CMS}

\newcommand{\mr}{\mathrm}

\newcommand{\PL}{Phys. Lett.}
\newcommand{\PR}{Phys. Rept.}

\newcommand{\EPJ}{Eur. Phys. J.}

\newcommand{\al}{{\it et al.}}

\begin{document}

\title{Forward physics with CMS}

\author{Marek Ta\v{s}evsk\'{y}}

\address{Physics Department of the Antwerp University\\
Universiteitsplein 1, B-2610 Antwerp, Belgium}

\maketitle

\abstracts{The physics potential of the forward physics project at CMS is very
rich. Some of the diffraction and low-$x$ physics channels are briefly
discussed.}

\vspace*{-5.76cm}
\begin{flushright}
CMS CR 2004/029
\end{flushright}
\vspace*{4.76cm}

\section{Introduction}
Data from the proton-proton collisions which will be produced at the LHC
at the highest-ever centre-of-mass energy, $\sqrt s$, of 14 TeV
will provide information about unexplored phase space regions and new
physics domains. The luminosity at startup, planned for the year 2007, is 
expected
to be $10^{33}$cm$^{-2}$s$^{-1}$. In the high luminosity mode, it will
reach values of $10^{34}$cm$^{-2}$s$^{-1}$ leading to an integrated luminosity
of 100 fb$^{-1}$ per year. At these highest luminosities, it is expected to see
on average 23 overlapping (mostly soft) hadronic interactions per bunch
crossing.

\begin{figure}[hbtp]
  \begin{center}
    \resizebox{12cm}{5cm}{\includegraphics{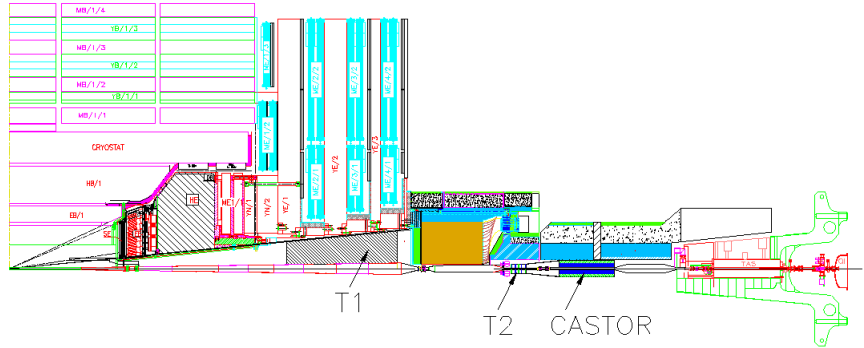}}
    \caption{Positions of the T1 and T2 TOTEM telescopes and the CASTOR
calorimeter, integrated into CMS.}
    \label{sketch}
  \end{center}
\end{figure}

The forward physics project includes the CMS \cite{CMS}, TOTEM \cite{TOTEM}
and CASTOR \cite{CASTOR} detectors.  The CMS detector is a general
purpose detector with an acceptance of $|\eta|<3$ for tracking and of
$|\eta|<5$ for calorimetry. The pseudorapidity is defined as
$\eta=-\ln\tan(\theta/2)$ where $\theta$ is the polar angle of a particle with
respect to the beam axis.
The TOTEM experiment will use the same interaction
point (IP) as CMS and is designed to measure the total and elastic pp cross
sections, and the diffraction dissociation. It will use two telescopes to
detect inelastic events, namely T1 with an acceptance of $3<|\eta|<5$ and T2
with an acceptance of $5.3<|\eta|<6.6$ (Fig.~\ref{sketch}), and three Roman Pot
(RP) stations, placed symmetrically (at 147, 180 and 215~m) from the IP to
measure protons scattered under very small $\theta$ angles.
The CASTOR calorimeter is designed to cover the same region as T2.
A combination of these detectors thus provides the largest acceptance
detector ever built at a hadron collider and enables a study of a large
variety of processes in detail. The issue of a
common usage of the TOTEM and CASTOR detectors an their integration into the
CMS detector and trigger/DAQ system is being investigated in a common study
group \cite{fwg} established in 2002.

\section{Diffraction}\label{diffr}
The selection of diffraction events is based on the presence of rapidity gaps
and non-dissociated protons. At the highest luminosities, it is impossible
to see any rapidity gaps due to overlap events and therefore, RP detectors are
needed to detect the scattered protons. At startup, still about
20\% of events will be of 'one-interaction-per-bunch-crossing' nature.
The low luminosity mode will therefore be useful for precise soft diffraction
measurements, such as those of different cross section components (single and
double dissociation, single and double pomeron exchange, ...).
The combination of the central and RP detectors enables the measurement of the
hard diffraction processes, such as single pomeron exchange (SPE) and
double pomeron exchange (DPE). These types of processes are useful to study
the pomeron structure and the dynamics of diffraction. The fractional momentum
of the pomeron carried by a parton entering the hard interaction is given by
$\beta = \Sigma_{\mr j\mr e\mr t\mr s} E_{\mr T} e^{-\eta}/(\sqrt {s}\cdot\xi)$, where the
jet characteristics are measured in CMS, while $\xi$, the proton momentum
loss, is measured in the RPs. The dynamics of diffraction is investigated by
studying the production of heavy particles, such as W and Z bosons, and heavy
quarks.

\begin{figure}[t]
  \begin{center}
    \resizebox{11cm}{10.5cm}{\includegraphics{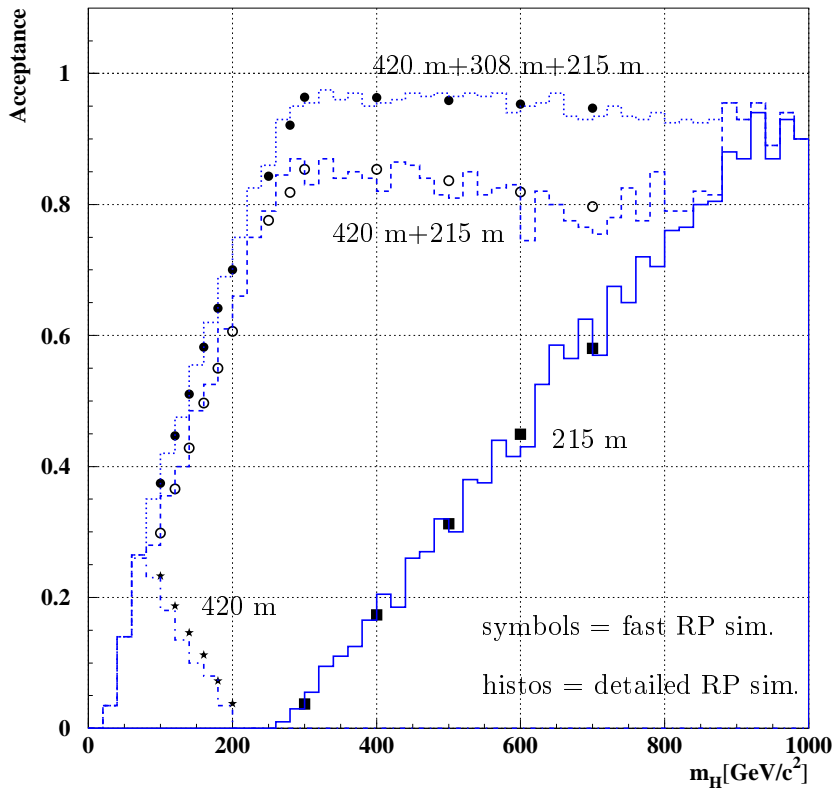}}
    \caption{Acceptance of various Roman Pot combinations as a function of the
Higgs boson mass. The histograms represent the results of Helsinki group study
[7] and the symbols represent the results of the CMS fast simulation using the
EDDE [8] event generator.}
    \label{acc}
  \end{center}
\end{figure}

The DPE processes with a Higgs boson in the final state are of particular
interest.
A recent calculation \cite{ways} for the DPE exclusive production of a
120~GeV/$c^2$ Higgs boson, pp$\rightarrow$ pHp, gives a cross section of about
3~fb, while that for the DPE inclusive production could be as large as
50--200~fb.
The exclusive Higgs boson production (the energy of the pomerons entirely goes into
the production of the Higgs boson, i.e. $\beta=1$) has the advantage of the
spin selection rule $J_Z=0$ \cite{jz0} which suppresses LO QCD background -
b$\bar{\mr b}$
production- allowing the H $\rightarrow$ b$\bar{\mr b}$ decay to be observed
in the central detector. Another advantage is a precise Higgs boson mass
determination by exploiting the missing mass method, $M^2_{\mr H} =
(p_1+p_2-p_3-p_4)^2$, where $p_1,p_2$ are the four-momenta of the beam protons
and $p_3,p_4$ those of the scattered protons measured in RPs.
First studies show that a mass resolution of 2--3\% can be achieved
\cite{helsinki}.
In Fig.~\ref{acc} the Higgs boson mass acceptance is shown for various
combinations of RP stations. A nice agreement is seen between the results of a
study of the RPs' response \cite{helsinki} and those of a fast simulation
program of CMS in which the former results were used as input together with the
EDDE \cite{EDDE} event generator. It turns out that the most distant RP
station of those designed by TOTEM so far (at 215~m) does not suffice to detect
a Higgs boson with a mass around 120~GeV/$c^2$. For that purpose, an additional
RP station further away from the IP would be needed. Two positions were
considered in \cite{helsinki}, namely 308 and 420~m. The combination of the
215~m and 420~m RP stations gives a 40\% acceptance for
$m_{\mr H}=120$~GeV/$c^2$. Anything
behind 215~m would, however, be in the so called ``cold region'', which means
difficulties in controlling and maintaining the stations. Furthermore, signals
from these distant stations would arrive too late to the central trigger to be
included in its first level. These issues are now intensively studied in the
common CMS/TOTEM working group.

\section{The low-$x$ programme}\label{lowx}
Data from HERA convincingly show a rise of the proton parton density function
(PDF) for $x$ down to $10^{-4}$. The PDF behaviour at even smaller $x$ is a
subject of intensive discussions. An important question is whether the PDF has 
reached
a region of saturation. In such a region, effects such as parton recombination
and shadowing corrections would suppress the growth of the PDF. At LHC,
$x$ values down to $10^{-7}$ can be reached \cite{martin}. Processes suitable
to extract the very low-$x$ region of the PDF are the production of low-mass
Drell-Yan lepton pair, prompt photons, W and dijets. The experimental
signature of the Drell-Yan events consists of a low-mass lepton pair going
very forward. The parton $x_1$ and $x_2$ values and the invariant mass of
the dimuon system, $M_{\mu\mu}$, are connected by the relation
$M_{\mu\mu}^2\sim x_1x_2s$. Hence, e.g. $x_1$ can be probed at very small
values if $M_{\mu\mu}$ is kept small and $x_2$ large ($x_2>0.1$). It has been
shown \cite{albert} that such a lepton pair predominantly goes in the very
forward direction.
Very-forward high $p_T$ leptons, photons and jets are expected
to be well measured in the CASTOR calorimeter.

\begin{figure}[hbtp]
  \begin{center}
    \resizebox{5.5cm}{7.5cm}{\includegraphics{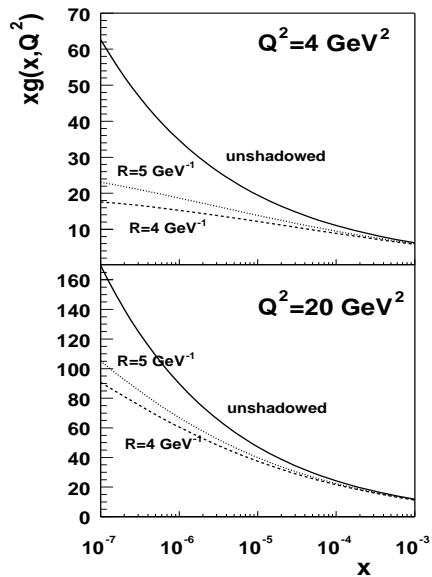}}
    \caption{Predictions for the gluon PDF for two hard scales $Q^2$, with and
without shadowing corrections. The shadowing corrections are shown for two
saturation radii $R$.}
    \label{shadow}
  \end{center}
\end{figure}

The effect of shadowing corrections for different saturation radii is shown
in Fig.~\ref{shadow}. The shadowing corrections were estimated with GLR type
of corrections to the standard evolution equations, using the results from
triple pomeron vertex calculations \cite{shad}. The effect turns out to be
sizeable (a factor of two) for $x=10^{-6}$ at $Q^2=4$~GeV$^2$ and is reduced
for $Q^2=20$~GeV$^2$ to values of 20--30\%.

\section{Other topics}
Other physics chapters included in the forward physics project are as follows
\cite{EOI}.\vspace*{0.2cm}\\
{\bf Two-Photon Physics}: The effective luminosity of high-energy
$\gamma\gamma$ collisions reaches 1\% of the pp luminosity and RP stations
should allow a reliable detection of scattered protons, hence of
$\gamma\gamma$ events. Possible measurements are the total $\gamma\gamma$
cross section, comparisons with QCD predictions, exclusive Higgs boson
production and Supersymmetry particle production \cite{Krystof}.
\vspace*{0.2cm}\\
{\bf Cosmic rays}: Programs used to reconstruct the incident particle energy
and type show large uncertainties, in particular in the forward direction.
Therefore, there is a considerable interest from the cosmic rays community
to see measurements of the particle and energy flow at large rapidities in
pp and pA interactions \cite{cosmic}. \vspace*{0.2cm} \\
{\bf Luminosity}: In the QED processes pp $\rightarrow$ ppe$^+$e$^-$, the
produced electrons point dominantly to the region $5<|\eta|<8$, which is in
the T2/CASTOR acceptance. Calculations can be controlled to a precision of
1\%. This channel is therefore promising for an absolute luminosity
measurement.

\section{Acknowledgements}
I would like to thank Patrick Janot for a careful reading of the manuscript.
The work was supported by Interuniversity Attraction Poles Programme -
Belgian Science Policy.

\end{document}